\documentclass[english,twocolumn,showpacs,preprintnumbers,amsmath,prd,amssymb,superscriptaddress]{revtex4}

\pdfoutput=1

\usepackage{graphicx}
\usepackage{dcolumn}
\usepackage{bm}

\usepackage{graphicx}
\usepackage{color}

\def\nuebar{\bar{\nu_{\rm e}}}
\def\numq{\delta_{\rm Q}}
\def\munu{\mu_{\nu}}

\newcommand{\be}{\begin{eqnarray}}
\newcommand{\ee}{\end{eqnarray}}

\usepackage{babel}
\makeatother

\begin{document}

\title{
Constraints on Millicharged Neutrinos via
Analysis of Data from Atomic Ionizations 
with Germanium Detectors at sub-keV Sensitivities
}

\author{Jiunn-Wei~Chen}
\affiliation{Department of Physics, National Taiwan University, 
Taipei 10617, Taiwan}
\affiliation{National Center for Theoretical Sciences 
and Leung Center for Cosmology and Particle Astrophysics, 
National Taiwan University, Taipei 10617, Taiwan}

\author{Hsin-Chang~Chi}
\affiliation{Department of Physics, National Dong Hwa University, 
Shoufeng, Hualien 97401, Taiwan}

\author{Hau-Bin~Li}
\affiliation{Institute of Physics, Academia Sinica, Taipei 11529, Taiwan}

\author{C.-P.~Liu}
\affiliation{Department of Physics, National Dong Hwa University, 
Shoufeng, Hualien 97401, Taiwan}

\author{Lakhwinder~Singh}
\altaffiliation[Corresponding Author: ]{ lakhwinder@phys.sinica.edu.tw } 
\affiliation{Institute of Physics, Academia Sinica, Taipei 11529, Taiwan}
\affiliation{Department of Physics, Banaras Hindu University, 
Varanasi 221005, India}

\author{Henry~T.~Wong}
\altaffiliation[Corresponding Author: ]{ htwong@phys.sinica.edu.tw } 
\affiliation{Institute of Physics, Academia Sinica, Taipei 11529, Taiwan}

\author{Chih-Liang~Wu}
\altaffiliation[Corresponding Author: ]{ b97b02002@ntu.edu.tw } 
\affiliation{Department of Physics, National Taiwan University, 
Taipei 10617, Taiwan}
\affiliation{Institute of Physics, Academia Sinica, Taipei 11529, Taiwan}

\author{Chih-Pan~Wu}
\affiliation{Department of Physics, National Taiwan University, 
Taipei 10617, Taiwan}

\date{\today}

\begin{abstract}

With the advent of detectors with sub-keV sensitivities,
atomic ionization 
has been identified as a promising avenue to 
probe possible neutrino electromagnetic properties.
The interaction cross-sections
induced by millicharged neutrinos 
are evaluated with the {\it ab-initio} 
multi-configuration relativistic 
random-phase approximation.
There is significant enhancement at 
atomic binding energies compared to that when 
the electrons are taken as free particles. 
Positive signals would distinctly manifest as peaks 
at specific energies with known intensity ratios.
Selected reactor neutrino data 
with germanium detectors 
at analysis threshold as low as 300~eV are studied.
No such signatures are observed, and a
combined limit on the neutrino charge fraction of 
$| \numq | < 1.0 \times 10^{-12}$ at 90$\%$ confidence level
is derived.
\end{abstract}

\pacs{
14.60.Lm, 
13.15.+g, 
13.40.Gp 	
}
\keywords{
Neutrino Properties, 
Neutrino Interactions,
Electromagnetic form factors
}

\maketitle

%

The physical origin and experimental consequences
of finite neutrino masses and mixings~\cite{nu-pdg}
are not fully understood.
Investigations on anomalous 
neutrino properties and interactions~\cite{nuprop-pdg}
are crucial to address these fundamental questions and
may provide hints or constraints to
new physics beyond the Standard Model (SM).
An avenue is on the studies
of possible neutrino electromagnetic
interactions~\cite{nuprop-pdg,nuem-old,nuem-review}
which, in addition, offer the potentials to
differentiate between Majorana and Dirac neutrinos.
The neutrino electromagnetic form factors in
C, P and T-conserving theories can be
formulated as:
\begin{eqnarray}
\label{eq::nuemff}
\Gamma_{\rm em}^{\mu} ~~~ & \equiv &
F_1 \cdot \gamma^{\mu} ~ + ~ 
F_2 \cdot \sigma^{\mu \nu} \cdot q_{\nu} ~~ , ~~ {\rm with} \\
F_1 & = & 
\numq \cdot e_0 ~ + ~ \frac{1}{6} \cdot q^2 
\cdot \langle r_{\nu}^2 \rangle 
~~ , ~~ {\rm and} \\
F_2 & = & 
( - i ) \cdot \frac{\munu}{2 \cdot m_e} ~~ , ~~ {\rm where}
\end{eqnarray}
$\gamma^{\mu}$ and $\sigma^{\mu \nu}$ are the
standard QED matrices, $e_0$ and $m_e$ are the electron charge 
and mass, respectively, $q = ( q_0 , \vec{q} )$ is the four-momentum transfer,
while the neutrino properties are parametrized by
the neutrino fractional charge relative to the electron 
($\numq$ $-$ commonly referred to as ``neutrino millicharge'' 
in the literature),
the neutrino charge radius ($\langle r_{\nu}^2 \rangle$), and the 
anomalous neutrino magnetic 
moment ($\munu$)~\cite{nuem-old,nuem-review} 
in units of the Bohr magneton $\mu_{\rm B}$.
The $F_1$ and $F_2$ terms characterize 
neutrino interactions without and with a change of 
the helicity states, respectively.
The studies of $\numq$ and
$\langle r_{\nu}^2 \rangle$ should in general 
be coupled to those
due to SM-electroweak interactions 
to account for the possible interference effects among them.
For completeness, we note that two additional form factors
are possible~\cite{nuem-review}:
the electric dipole moments in theories violating
both P- and T-symmetries, and  
the anapole moments in P-violating theories.

\begin{figure}
{\bf (a)}\\
\includegraphics[width=8.0cm]{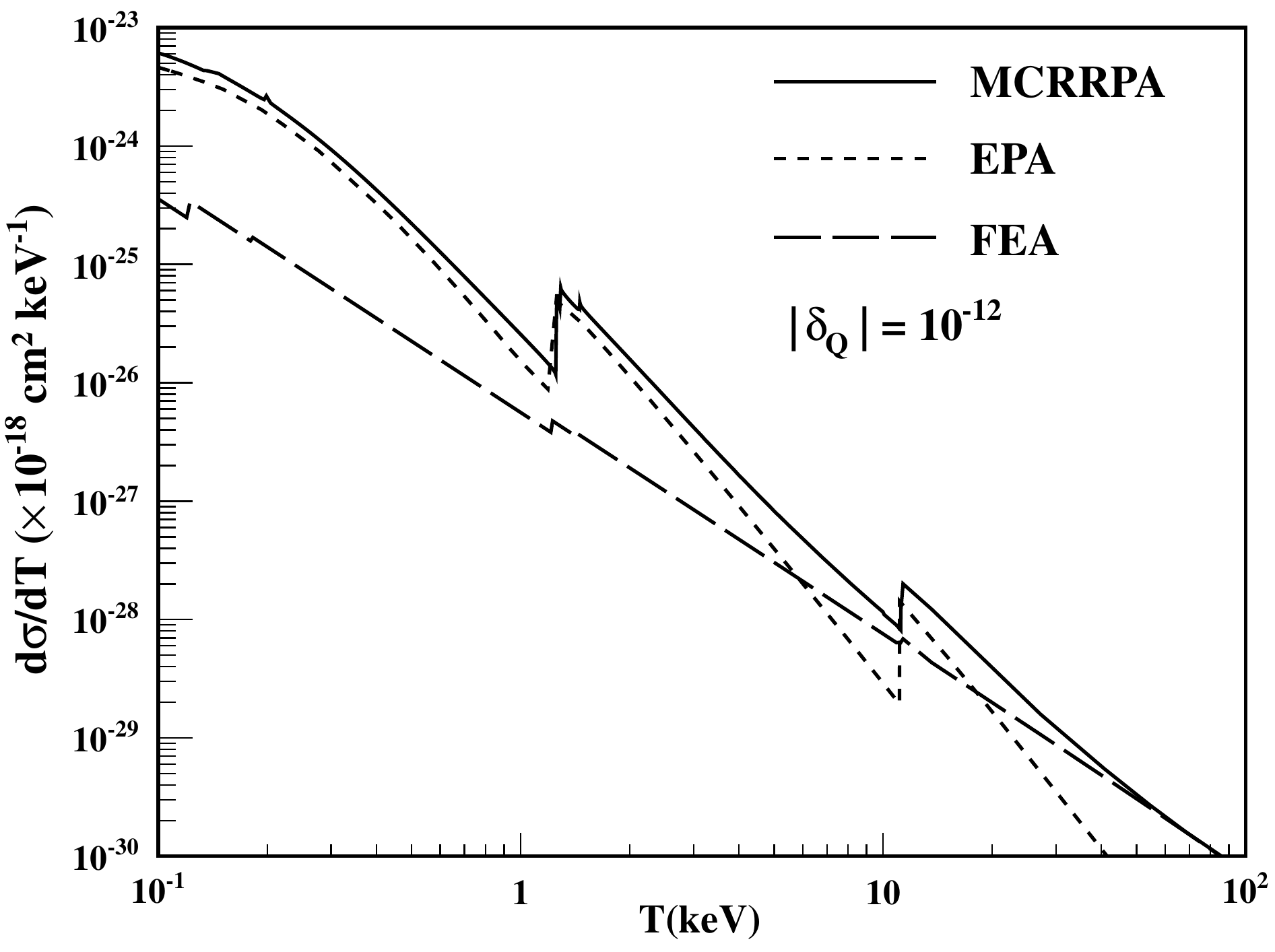}\\
{\bf (b)}\\
\includegraphics[width=8.0cm]{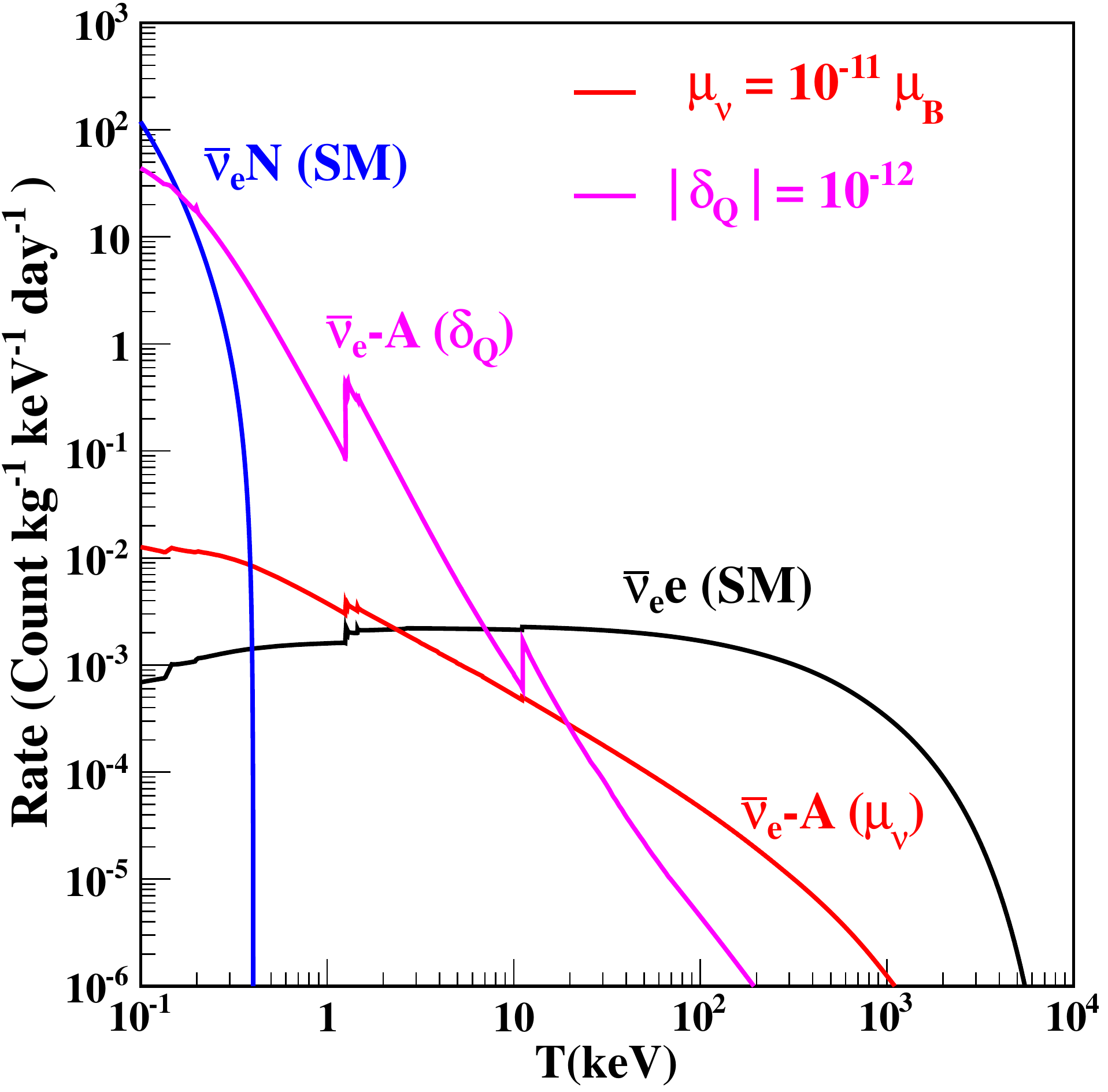}
\caption{
The differential cross-sections
$\nuebar$-A($\numq$) on Ge 
due to $| \numq | = 10^{-12}$ $-$
(a) with mono-chromatic $E_{\nu} = 1 ~ {\rm MeV}$,
derived with the FEA, EPA and MCRRPA methods,
and
(b)
 with typical reactor spectrum at a flux of
$\phi ( \nuebar ) = \rm{10^{13} ~ cm^{-2} s^{-1}}$, 
where contributions
from SM $\nuebar$-e and $\nuebar$-nucleus(N) 
as well as those of $\munu = 10^{-11} ~ \mu_{\rm B}$ 
are overlaid.
Standard quenching factors~\cite{texonopcge} 
are used to account for observable ionizations 
in nuclear recoils.
The contribution to $\nuebar$-e from
$\langle r_{\nu}^2 \rangle$ at its current upper bound
is only a fraction of that from SM, and is
not shown. 
}
\label{fig::dsigmadT}
\end{figure}

The theme of this article is to report
a new direct laboratory limit on $| \numq |$.
The searches are based on $\nuebar$
emitted from the nuclear power reactor 
via atomic ionization~\cite{nuemai},
an interaction channel considered for the first time in this process.
The cross-section is derived using the 
Multi-Configuration Relativistic Random-Phase
Approximation (MCRRPA) theory~\cite{mcrrpa,rpaplb14}.
As will be demonstrated in Figure~\ref{fig::dsigmadT}b,
the bounds on event rates from 
$\numq$-induced atomic interactions [$\nuebar$-A($\numq$)]:
\begin{equation}
\nuebar ~ + ~ {\rm A } ~ \rightarrow ~ \nuebar ~ 
+ ~ {\rm A^+ } ~ + ~ e^- ~
\end{equation}
to be probed in this work 
(${\rm \sim 1 ~ count~ kg^{-1} keV^{-1} day^{-1}}$ at
an energy transfer of $T \sim 0.1 - 10 ~ {\rm keV}$)
far exceed those due to SM interactions
as well as  $\langle r_{\nu}^2 \rangle$-induced
processes at its current limits~\cite{texononuecs},
such that these effects and their interference
can be neglected in our analysis.


The origin of electric charge quantization and whether it is exact
is one of Nature's profound mysteries. 
Many theories~\cite{QQ-theories}, 
such as extra dimensions, magnetic monopoles, 
and grand unified theories, 
provide elegant solutions, but they remain speculative.
Electric charges are quantized in SM due to 
{\it U(1)} gauge invariance
and anomaly cancellation~\cite{QQ-SM,DiracMajo}, implying $\numq = 0$.
However, charge quantization is no longer ensured
in many extensions of SM~\cite{DiracMajo,mQ-BSM}.
For example, in theories with right-handed neutrinos and 
Dirac mass terms, 
electric charge is no longer quantized and $\delta_Q$ 
can assume an arbitrary value due to 
a hidden $U(1)$ symmetry whose conserved charge 
is the difference of baryon and lepton numbers or $B-L$. 
Charge quantization can be restored
by introducing additional conditions such as Majorana mass terms 
that break the $U(1)$ symmetry~\cite{DiracMajo}. 
Neutrinos with finite charge will necessarily imply they are
Dirac particles. 


Model-dependent astrophysics bounds~\cite{nuprop-pdg,raffelt99}
ranging at $| \numq | < 10^{-13}$$-$$10^{-15}$ are
derived from stellar luminosity and cooling, as well as 
the absence of anomalous timing dispersion of 
the neutrino events in SN 1987A.
The most stringent indirect limit is 
$| \numq | < 3 \times 10^{-21}$~\cite{raffelt99},
inferred from constraints on the neutrality of 
the hydrogen atoms and the neutrons~\cite{neutrality},
and assuming charge conservation in neutron beta-decay.
Earlier efforts with direct laboratory experiments
placed constraints 
$| \numq | < {\rm few} \times 10^{-12}$~\cite{ethz,msu} 
through the extrapolations of the $\munu$-results from
reactor $\nuebar$ experiments~\cite{texonomunu,gemma}
using simplistic scaling relationships and 
neglecting atomic effects. 


\begin{table*}
\begin{ruledtabular}
\begin{tabular}{lcccccc}
Data Set & Reactor-$\nuebar$ & Data Strength & Analysis & 
\multicolumn{3}{c}{$| \numq |$ 90\% CL Limits ($< \times 10^{-12}$)}  \\
& Flux &  Reactor ON/OFF & Threshold &  Previous Analysis &
\multicolumn{2}{c}{This Work} \\ 
& (${\rm \times 10^{13} ~  cm^{-2} s^{-1}}$) & (kg-days) & (keV) 
& FEA & FEA & MCRRPA\\ \hline
TEXONO 1~kg Ge~\cite{texonomunu}  &   0.64 &
570.7/127.8   & 12    & 3.7~\cite{ethz} & 14 & 8.8   \\
GEMMA 1.5~kg Ge~\cite{gemma} & 2.7 & 1133.4/280.4 
& 2.8 & 1.5~\cite{msu} & 2.1 &  1.1 \\
TEXONO Point-Contact Ge~\cite{texonopcge} &  0.64 
& 124.2/70.3  & 0.3  & $-$ & $-$ & 2.1   \\
Projected Point-Contact Ge & 2.7 & 800/200  & 0.1 & $-$ 
& $-$ & $\sim$0.06 
\end{tabular}
\end{ruledtabular}
\caption{
Summary of experimental 
limits on millicharged neutrino at 90\% CL
with selected reactor neutrino data. 
``This Work'' compares data with 
results from FEA and MCRRPA calculations 
via a complete analysis,
while ``Previous Analysis'' is based on 
extrapolations from $\mu_{\nu}$-results using 
simplistic scaling relations to the FEA spectra.
The projected sensitivity of measurements
at the specified experimental parameters is also shown. 
}
\label{tab::results}
\end{table*}



The conventional way of evaluating the effects due to 
$\Gamma_{\rm em}^{\mu}$ is with the Free Electron Approximation (FEA).
The corresponding differential cross-section for
$\numq$-induced neutrino$-$electron scattering~\cite{ethz,msu} 
due to an incoming neutrino of energy $E_{\nu}$ 
at $T \ll E_{\nu}$ is:
\begin{equation}
\left( \frac{d \sigma}{dT} \right) _{\rm FEA} ~ = ~
\numq^2 ~
\left[ \frac{2  \pi  \alpha_{em}^2 }{m_e} \right] ~
\left[ \frac{1}{T^2} \right] ~~ ,
\label{eq::fea}
\end{equation}
where $\alpha_{em}$ is the fine structure constant. 
The $( 1/T^2 ) $-dependence is different from that of
$( 1/ T )$ for $\munu$.
With the advent of low-energy detectors sensitive to 
the energy range of atomic transitions and binding energies 
($T < 10~{\rm keV}$),
FEA is no longer adequate and 
atomic ionization effects have to be taken into account~\cite{nuemai}.
Cross-sections of $\munu$-induced $\nu$-atom scattering
have been formulated~\cite{rpaplb14,munuaitheory,twprd13} 
by various authors.


The cross-section $\nuebar$-A($\numq$)
is analogous to that induced by
relativistic charged leptons, and can be
described at atomic energies 
by the Equivalent Photon Approximation (EPA)~\cite{epa}:
\begin{equation}
\left( \frac{d \sigma}{dT} \right) _{\rm EPA} ~ =  ~ 
\numq^2 ~
\left[ \frac{2 ~ \alpha_{em}}{\pi} \right] ~
\left[ \frac{\sigma_{\gamma}(T)}{T} \right] ~
{\rm log} [ \frac{E_{\nu}}{m_{\nu}} ] ~~ ,
\label{eq::epa}
\end{equation}
where $m_{\nu}$ is the neutrino
and $\sigma_{\gamma} ( T )$ is 
the photo-electric cross-section
by a real photon of energy $T$~\cite{Gephoto}.
The divergence at $m_{\nu} \rightarrow 0$ is expected 
in Coulomb scattering and some cutoff schemes are necessary.
The Debye length for solid-Ge (0.68~$\mu$m or 0.29~eV), 
which characterizes the scale of screen Coulomb interaction, 
is chosen. This may introduce an uncertainty of $\sim$20\% 
to the normalization if $m_{\nu}$ would be replaced
by other values related to neutrino mass bounds.
The EPA method neglects the contributions from the
longitudinal polarization of the virtual photons
and hence would deviate from the correct results
as $T$ increases.
It fails to describe ionizations
by $\munu$, $\langle r_{\nu}^2 \rangle$ or
electro-weak interactions~\cite{twprd13}.


We adopted the MCRRPA theory~\cite{mcrrpa}
as an {\it ab-initio} approach~\cite{rpaplb14}
to provide an improved description of 
the atomic many-body effects.
This becomes relevant 
for data from Ge detectors at sub-keV sensitivities.
The MCRRPA theory is a generalization of relativistic 
random-phase approximation (RRPA)
by the use of a multi-configuration wave function 
as the reference state. 
It has been successfully applied 
to photo-excitation, photo-ionization and 
$\munu$-induced ionization of divalent or 
quasi-divalent atomic systems~\cite{mcrrpa,rpaplb14}.
There are various aspects where 
MCRRPA improves over the time-dependent Hartree–Fock (HF) approximation
in describing the structures and transitions of Ge $-$
(i) The Ge atom has two valence $4p$ electrons. 
Its ground state, a $^3 P_0$ state, can be formed by either a
$4 p^2_{\frac{1}{2}}$ or a $4 p^2_{\frac{3}{2}}$ valence configuration. 
This entails the necessity of a 
multi-configuration reference state. 
(ii) With an atomic number of $Z$=32, the relativistic 
corrections, in power of $Z \alpha_{em}$$\sim$1/4, 
can no longer be ignored. 
By solving a relativistic wave equation, the leading 
relativistic effects
are included non-perturbatively from the onset. 
(iii) The two-body correlation beyond HF is generally important 
in building excited states. The RRPA is
an established method in accounting for two-body correlation, 
having nice features such as treating the reference
and excited states on the same footing and preserving gauge invariance. 
In combination with the multi-configuration reference state, 
configuration mixing due to two-body correlation is also 
taken into account.


\begin{figure}
\includegraphics[width=8.0cm]{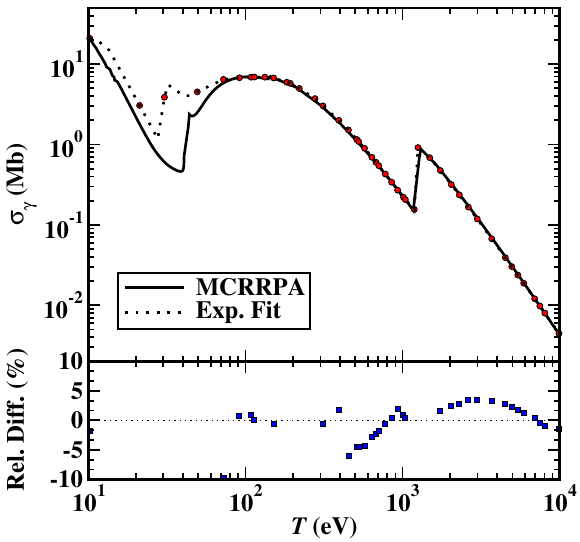}
\caption{
Germanium photo-ionization cross section. The solid curve
corresponds to the results of MCRRPA calculation.
The dotted line is a fit to experimental data 
taken from Ref.~\cite{Gephoto}.
The relative differences (excess of MCRRPA results over data
relative to the measurements) are shown in the lower panel.
}
\label{fig::sigmaphoto}
\end{figure}


The MCRRPA results of this work are benchmarked 
by the measured photo-absorption cross section of solid Ge 
from real photons~\cite{rpaplb14,Gephoto}. 
As demonstrated in Figure~\ref{fig::sigmaphoto},
the calculations successfully reproduce the data
to an accuracy of within 5\% at energy transfer
larger than 100~eV, where 
the inner-shell electrons of Ge ($3p$ and below) 
provide the dominant contributions. 
The deviations originate from the small contributions of 
the outer-shell electrons. 
At lower energy,
the solid state effects start to play a role,
since Ge is fabricated as semiconductor crystals
in ionization detectors.


The derived differential cross-section for $\nuebar$-A($\numq$)
on Ge under various schemes are depicted in Figure~\ref{fig::dsigmadT}a 
with a mono-chromatic incident neutrino at 
$E_{\nu} = 1~ {\rm MeV}$, a typical range for reactor-$\nuebar$.
The FEA scheme is expected to provide good descriptions 
at energy transfer larger than the atomic binding energy scale,
while EPA at $q^2 \rightarrow 0$.
The MCRRPA results converge to these benchmarks: 
$ T > 50 ~ {\rm keV}$ for FEA and $T < 1~{\rm keV}$ for EPA, 
confirming the method covers a wide range of validity. 
Two features are particularly note-worthy:
(i) There is an order-of-magnitude enhancement 
in the MCRRPA or EPA
cross-section over FEA at low energy
when atomic effects are taken into account.
This behaviour is opposite to that for $\munu$-induced
interactions~\cite{rpaplb14,munuaitheory} where the
cross-section is suppressed, 
the origin of which is discussed in Ref.~\cite{twprd13}.
(ii) There exists a unique ``smoking gun'' signature
for $\nuebar$-A($\numq$), through the observation
of K- and L-shell peaks at the specific binding energies 
and with known intensity ratios.
Both features favor the use of detectors with low
threshold at sub-keV energy, and yet possess 
good resolution to resolve peaks and other structures
at such energy.

To be comparable with experimental data,
the differential cross-sections are convoluted with
the neutrino spectrum ${d \phi}/{dE_{\nu}}$ to provide
the observable spectrum of event rates ($R$) 
as function of $T$:
\begin{equation}
\frac{d R}{d T} ~ = ~ \rho_{e} 
 \int_{E_{\nu}}
\left[ \frac{d\sigma}{dT} \right] ~ 
\left[ \frac{d\phi}{dE_{\nu}} \right] ~
dE_{\nu} ~~ ,
\label{eq::expt-rate} 
\end{equation}
where, $\rho_{e}$ is the electron number density 
per unit target mass.
The MCRRPA spectrum for [$\nuebar$-A($\numq$)] with Ge
at a typical reactor neutrino flux of 
$\phi(\nuebar) = 10^{13} ~ {\rm cm^{-1} s^{-1}}$
is depicted in Figure~\ref{fig::dsigmadT}b, and is compared
with those of $\munu$-induced 
and SM $\nuebar$-e and 
$\nuebar$-nucleus coherent scatterings~\cite{nuNcohsca}.
It can be seen that low threshold detectors
can greatly enhance the sensitivities in most
of the channels.


\begin{figure}
\includegraphics[width=8.0cm]{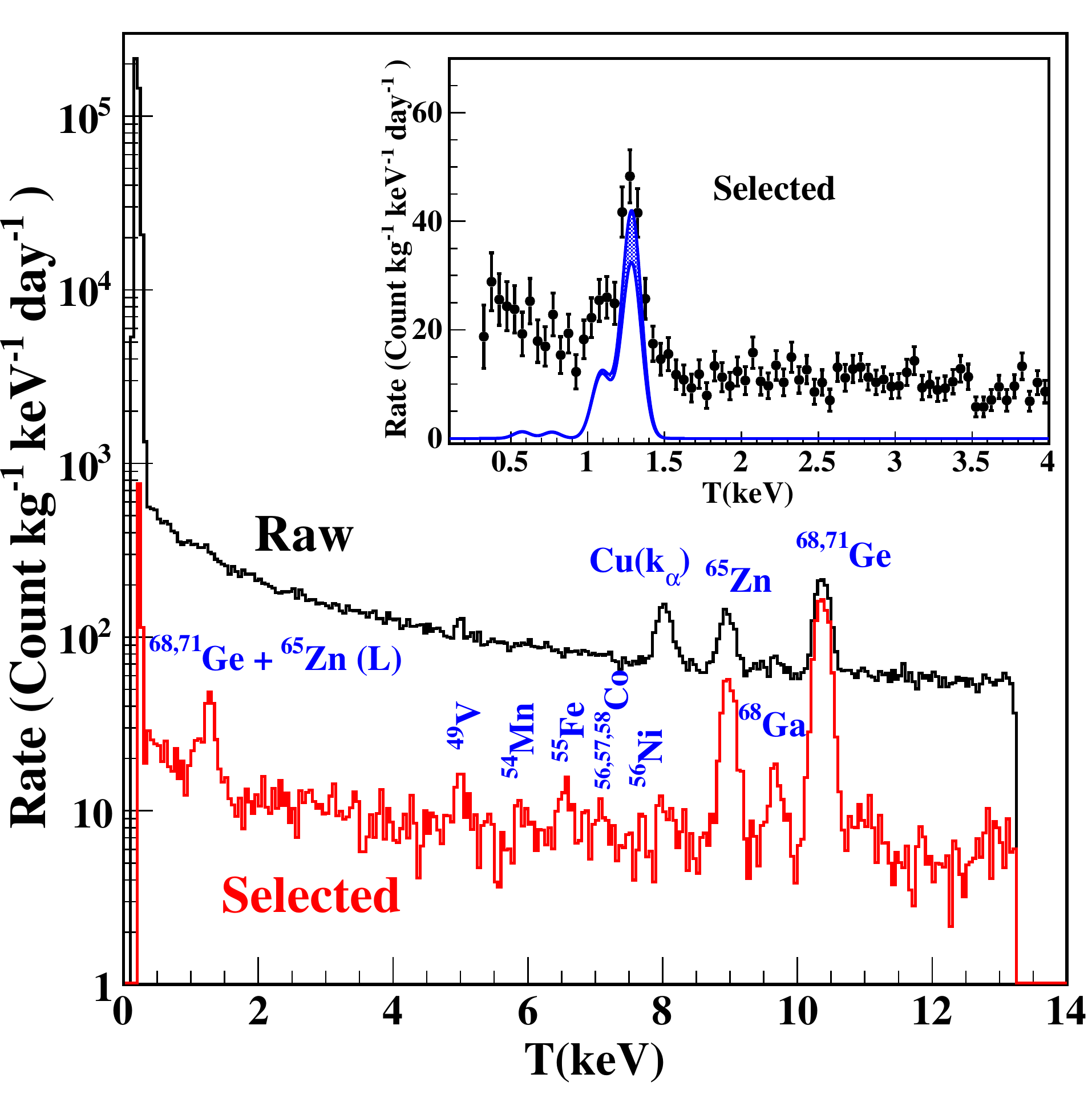}
\caption{
Typical spectra from point-contact Ge detectors at 
the Kuo-Sheng Reactor Neutrino Laboratory.
The ``Selected'' spectra are due 
to events after having anti-coincidence with the 
cosmic-ray and anti-Compton detectors.
The peaks are due to internal radioactivity.
The low energy spectrum is expanded in the inset. 
Intensities of the L-shell X-rays 
can be independently derived from 
the higher-energy K-shell peaks.
}
\label{fig::pcgespectra}
\end{figure}

\begin{figure}
\includegraphics[width=8.0cm]{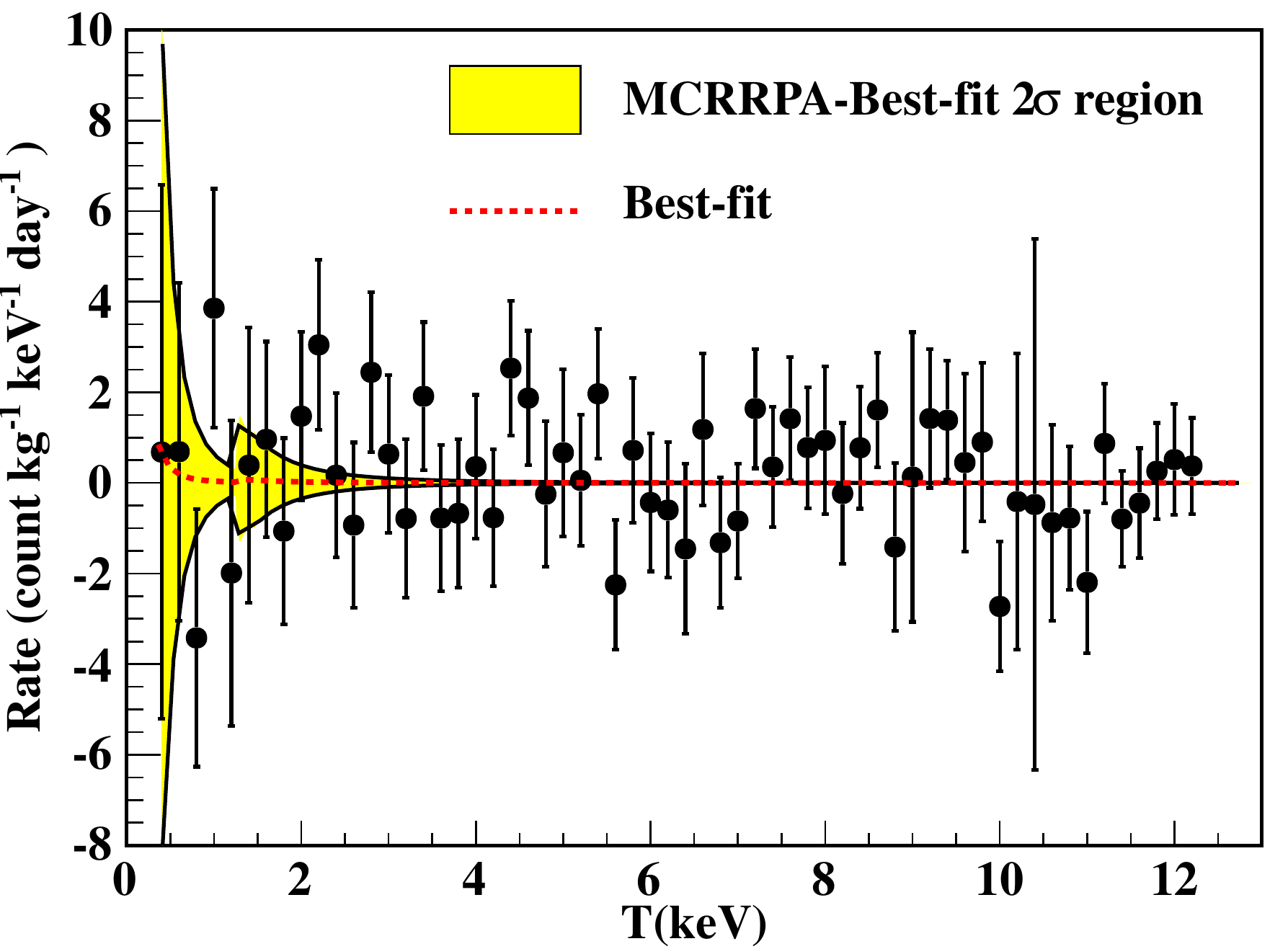}
\caption{
Analysis of $\nuebar$ millicharge with data
from point-contact Ge detectors~\cite{texonopcge},
showing Reactor ON$-$OFF spectrum
with 124.2(70.3)~kg-days of ON(OFF) data.
The best-fit function with its 2$\sigma$ uncertainty band 
is superimposed.
}
\label{fig::pcgeanalysis}
\end{figure}

The previous analysis~\cite{ethz,msu} with FEA
are repeated using full spectral data via
standard statistical procedures.
Comparisons of the results listed in Table~\ref{tab::results}
show discrepancies and indicate inadequacies
of the scaling approach.
Also displayed are the analysis of 
reactor $\nuebar$ data~\cite{texonomunu,gemma,texonopcge}
using the MCRRPA spectrum of Figure~\ref{fig::dsigmadT}b.
No evidence on $\nuebar$-A($\numq$) are observed and 
limits of $| \numq |$ at 90\% confidence level (CL) are 
derived.
In particular, we illustrate the results from
point-contact Ge detectors
with sub-keV sensitivities at
analysis threshold of 300~eV~\cite{texonopcge}.
The detectors are deployed by the TEXONO experiment
at Kuo-Sheng Reactor Neutrino Laboratory~\cite{texonomunu,texononuecs}
for the studies of
light WIMP dark matter and $\nuebar$-nucleus 
coherent scatterings.
Typical spectra are depicted in Figure~\ref{fig::pcgespectra}. 
The ``Raw'' spectrum is due to all events prior to background
suppression, while the ``Selected'' ones are those
of candidate events having anti-coincidence with the 
cosmic-ray and anti-Compton detectors.
Various lines from internal X-rays emissions 
due to cosmic-induced internal radioactivity 
can be observed.
The low energy portion of the candidate spectrum 
is displayed in the inset. 
The peaks are from the L-shell X-rays,  
where the intensities can be quantitatively accounted for 
with the higher-energy K-shell peaks.
Data taken in different reactor periods are combined
and compared with $\nuebar$-A($\numq$) from MCRRPA.
Depicted in Figure~\ref{fig::pcgeanalysis}
is reactor ON$-$OFF residual spectrum 
with 124.2(70.3)~kg-days of ON(OFF) data.
The best-fit solution with the 2$\sigma$ uncertainty
band is superimposed. 
A limit of $| \numq | < 2.1 \times 10^{-12}$  
at 90\% CL is derived.

The various bounds derived from the MCRRPA analysis
shown in Table~\ref{tab::results} can be statistically
combined.
The overall limit 
from these reactor neutrino data is  
\begin{equation}
| \numq | ~ < ~ 1.0 \times 10^{-12}   ~~~ 
\end{equation}
at 90\% CL. 
The projected sensitivity for a
measurement at an achieved flux and data strength~\cite{gemma}
together with 100~eV detector threshold targeted for
next generation of experiments~\cite{nuNcohsca}
would be $| \numq | \sim 6 \times 10^{-14} $. 


The MCRRPA theory improves descriptions
on neutrino electromagnetic effects at
atomic energy scales over previous techniques. 
Possible charge-induced interactions
show enhancement at atomic binding energies, and
would manifest as peaks with known intensity ratios.
Novel Ge-detectors with sub-keV sensitivities
and superb energy resolution are ideal to study 
these effects.
We plan to extend our studies to neutrinos at
different kinematics regimes, as well as to
possible electromagnetic interactions with WIMPs
as non-relativistic particles.

This work is supported by
the Academia Sinica Investigator Award 2011-15 (HTW),
as well as 
contracts 102-2112-M-001-018 (LS,HBL),
102-2112-M-002-013-MY3 (JWC,CLW,CPW),
102-2112-M-259-005 (CPL),
from the Ministry of Science and Technology, Taiwan.

\end{document}